\documentclass[a4paper,12pt]{article}

\usepackage{amsmath}
\usepackage[psamsfonts]{amssymb}
\usepackage{rsfs}
\usepackage{bm}

\usepackage{cite}
\usepackage[dvips]{graphicx}
\usepackage{color}

\makeatletter
\@addtoreset{equation}{section}
\makeatother

\begin{document} 

\begin{titlepage}

\baselineskip 10pt
\vskip 5pt
\leftline{}
\leftline{KEK/Chiba Univ. Preprint
          \hfill   \small \hbox{\bf KEK Preprint 2009-32/ CHIBA-EP-181}}
\leftline{\hfill   \small \hbox{November 2009}}
\vskip 5pt
\baselineskip 14pt
\centerline{\Large\bf 
} 
\vskip 0.5cm
\centerline{\Large\bf  
The exact decomposition of gauge variables  
}
\vskip 0.3cm
\centerline{\Large\bf  
in lattice Yang-Mills theory
}

\vskip 0.3cm

\centerline{{\bf 
Akihiro Shibata,$^{\flat,{1}}$ Kei-Ichi Kondo,$^{\dagger,\ddagger,\sharp,{2}}$  and Toru Shinohara$^{\ddagger,{3}}$
}}  
\vskip 0.3cm
\centerline{\it
${}^{\flat}$Computing Research Center, High Energy Accelerator Research Organization (KEK)   
}
\vskip 0.1cm
\centerline{\it
\& 
Graduate Univ. for Advanced Studies (Sokendai), 
Tsukuba 
305-0801, 
Japan
}
\vskip 0.3cm
\centerline{\it
${}^{\dagger}$Department of Physics, University of Tokyo,
Tokyo 113-0033, Japan
}
\vskip 0.3cm
\centerline{\it
${}^{\ddagger}$Department of Physics, 
Graduate School of Science, 
Chiba University, Chiba 263-8522, Japan
}
\vskip 0.3cm

\begin{abstract}
In this paper, we consider lattice versions of the decomposition of the  Yang-Mills field a la Cho-Faddeev-Niemi, which was extended by Kondo, Shinohara and Murakami in the continuum formulation. 
For the SU(N) gauge group, we propose a set of defining equations for specifying the decomposition of the gauge link variable and solve them exactly without using the  ansatz adopted in the previous studies for $SU(2)$ and $SU(3)$.  
As a result, we obtain the general form of the decomposition for $SU(N)$ gauge link variables and confirm the previous results obtained for $SU(2)$ and $SU(3)$. 

\end{abstract}

 

\hrule  
${}^\sharp$ 
On sabbatical leave of absence from Chiba University. 
\\
  E-mail:  
${}^1${\tt akihiro.shibata@kek.jp};

${}^2${\tt kondok@faculty.chiba-u.jp};
${}^3${\tt sinohara@graduate.chiba-u.jp};

\par 
\par\noindent


\vskip 0.3cm

\pagenumbering{roman}
\tableofcontents




\end{titlepage}


\pagenumbering{arabic}

\baselineskip 14pt

\section{Introduction}

If one regards the dual superconductivity \cite{dualsuper} as a promising scenario for understanding quark confinement, one has to show the existence of magnetic monopole in the Yang-Mills theory \cite{YM54}, which is an indispensable ingredient for causing the dual superconductivity.  
One can recall a few examples of magnetic monopoles defined in gauge field theories. 
 In the Maxwell electromagnetism,  the Dirac magnetic monopole is realized by introducing singularities in the gauge potential.  Otherwise, the Bianchi identity leads to identically vanishing magnetic current. 
 In the non-Abelian gauge theory with (adjoint) matter fields such as the Georgi-Glashow model,  one can construct the 't Hooft-Polyakov magnetic monopole without introducing the singularity in the Yang-Mills field thanks to the extra degrees of freedom of matter fields. 

In pure Yang-Mills theory in question, two methods are so far known in realizing magnetic monopoles even in the absence of matter fields: 
\begin{enumerate}

\item
  Abelian projection \cite{tHooft81}  and maximal Abelian gauge \cite{KLSW87} 

\item
  Decomposition of the Yang-Mills field variable \cite{DG79,Cho80,FN98,Shabanov99,Cho80c,FN99a} and change of variables \cite{KMS06,KSM08,Kondo08,KS08}
  

\end{enumerate}

The first method, i.e., Abelian projection (as a partial gauge fixing) proposed by 't Hooft \cite{tHooft81} was conventionally used to extract the magnetic monopole from the Yang-Mills field.  
In the maximal Abelian (MA) gauge \cite{KLSW87}, the (infrared) Abelian dominance \cite{EI82} was confirmed by numerical simulations on the lattice for the string tension \cite{SY90} and correlation functions \cite{AS99}, while  magnetic monopole dominance was also confirmed in the string tension \cite{SNW94}.   See \cite{review} for reviews. 

The second method based on the CFN decomposition of the Yang-Mills field a la Cho \cite{Cho80} and Faddeev-Niemi \cite{FN98} has been developed especially in the last decade.  The second method is recognized to be superior in some aspects to the first one:  The magnetic monopole can be constructed in a manifestly gauge-independent way. 
In particular, the MA gauge is reproduced as a special limit of the second method.  In other words, the first method is nothing but a gauge-fixed version of the second method.  

In view of these, we have investigated the lattice versions of the CFN decomposition in the previous papers \cite{KKMSSI05,IKKMSS06,SKKMSI07} for $SU(2)$ and \cite{SKKMSI07b,KSSMKI08} for $SU(3)$, which enable us to perform numerical simulations.   
We have given the explicit forms for the new lattice variables $V_{x,\mu}$ and $X_{x,\mu}$ in terms of the original link variable $U_{x,\mu}$ and the color field $\bm{n}_{x}$. 
They are obtained by solving the defining equations which  are coupled  matrix equations. 
In order to solve them, in practice, we have so far assumed an ansatz written in terms of $U_{x,\mu}$ and $\bm{n}_{x}$  and determined the parameters in the ansatz by using the defining equations. The result was justified from the coincidence with the continuum expression in the naive continuum limit. 
Therefore, this procedure does not guarantee the generality or uniqueness of the solution for arbitrary lattice spacing.

In this paper, we propose the defining equation of the lattice CFN decomposition and solve them without using any ansatz to obtain the general and exact solution for $SU(N)$ on the lattice with arbitrary lattice spacing. 
Remarkably, the resulting expressions for $SU(2)$ and $SU(3)$ have the same form as those obtained previously  \cite{KKMSSI05,IKKMSS06,SKKMSI07,KSSMKI08,SKKMSI07b}. 
Therefore, the result of this paper confirms the generality and uniqueness of the previous solutions, in addition to the general solution in the case of $SU(N), N>3$ for the maximal and minimal options. 
In particular, the CFN decomposition given in this paper is intrinsic on a lattice without reference to the continuum limit.

\section{Defining equation of lattice CFN decomposition}

We adopt the $D$-dimensional Euclidean lattice $L_\epsilon=(\epsilon \mathbb{Z})^{D}$ with a lattice spacing $\epsilon$. 
In the lattice gauge theory with a gauge group $G$, the gauge variable $U_\ell=U_{x,\mu}$ is defined on an oriented link $\ell=<x,x+\epsilon \mu> \in L_\epsilon$ running from $x$ to $x+\epsilon \mu$ as
\footnote{In order to consider the naive continuum limit, the correspondence between the Lie group $U$ and the Lie algebra $\mathscr{A}$ is given by the mid-point definition:
$ 
 U_{x,\mu} =    \exp \left\{ -ig \epsilon \mathscr{A}_\mu(x^\prime) \right\} 
$ 
where $x^\prime:=x+\epsilon \mu/2$ is the midpoint of the link $<x,x+\epsilon \mu>$.
This prescription is adopted to suppress as much as possible lattice artifacts coming from a finite (nonzero) lattice spacing, in contrast to the very naive definition: $U_{x,\mu} =    \exp \left\{ -ig \epsilon \mathscr{A}_\mu(x) \right\}$.
}
\begin{equation}
 U_{x,\mu} =  \mathscr{P} \exp \left\{ -ig \int_{x}^{x+\epsilon \mu} dx^\mu \mathscr{A}_\mu(x) \right\} \in G.
\end{equation}
The link variable $U_{x,\mu}$ obeys the well-known lattice gauge transformation:
\begin{equation}
  U_{x,\mu}  \rightarrow \Omega_{x} U_{x,\mu} \Omega_{x+\mu}^{-1} = U_{x,\mu}^\prime
  , \quad \Omega_{x} \in G
  \label{U-transf}
 .
\end{equation}

In order to construct the lattice version of the CFN decomposition, we need to introduce the color field $\bm{n}_{x}$ which plays the crucial role in the CFN decomposition.
The color field is defined as an element of the coset space $G/\tilde{H}$ with $\tilde{H}$ being the stability group \cite{KSM08}: 
\begin{equation}
   \bm{n}_{x} \in G/\tilde{H}  .
\end{equation}
In case of $SU(2)$, the stability group $\tilde{H}$ is unique, i.e., a compact $U(1)$ group. While for $SU(3)$, there are two stability groups, i.e., $\tilde{H}=U(1)\times U(1)$ and $\tilde{H}=U(2)$. For $G=SU(N)$ ($N\geq 4)$, there exist more than $N-1$ stability groups. 
In particular, the maximal option corresponds to $\tilde{H}=U(1)^{N-1}$, while the minimal one  to  $\tilde{H}=U(N-1)$.

The color field $\bm{n}_{x}$ on a lattice is regarded as a site variable defined on a site $x$ and transforms in the adjoint way by another (independent) gauge rotation:
\begin{equation}
  \bm{n}_{x}  \rightarrow \Theta_{x} \bm{n}_{x} \Theta_{x}^{-1} = \bm{n}_{x}^\prime
  , \quad \Theta_{x} \in G/\tilde{H}
  .
  \label{n-transf1}
\end{equation}
By applying the reduction condition \cite{KMS06,KSM08},   the color field must transforms as 
\begin{equation}
  \bm{n}_{x}  \rightarrow \Omega_{x} \bm{n}_{x} \Omega_{x}^{-1} = \bm{n}_{x}^\prime
  , \quad \Omega_{x} \in G
  .
  \label{n-transf}
\end{equation}
In this paper, we do not discuss the reduction condition.  
See \cite{KKMSSI05,IKKMSS06,SKKMSI07,SKKMSI07b,KSSMKI08} for the reduction condition on the lattice.

For a given color field $\bm{n}_{x}$, we consider decomposing the $G$-valued gauge variable $U_\ell=U_{x,\mu} \in G$ into the product of two $G$-valued variables $X_{x,\mu}$ and $V_{x,\mu}$ defined on the same lattice \cite{KSSMKI08}:
\begin{equation}
 U_{x,\mu} = X_{x,\mu} V_{x,\mu} \in G ,
 \quad X_{x,\mu}, V_{x,\mu} \in G
  .
  \label{decomp-1}
\end{equation}
Here we require that $V_{x,\mu}$ is a new link variable which transforms like a usual gauge variable on the same  link $\ell=<x,x+\epsilon \mu>$: 
\begin{equation}
  V_{x,\mu}  \rightarrow \Omega_{x} V_{x,\mu} \Omega_{x+\mu}^{-1} = V_{x,\mu}^\prime 
  , \quad \Omega_{x} \in G
   .
   \label{V-transf}
\end{equation}
For this gauge transformation to be consistent with the decomposition (\ref{decomp-1}), consequently, $X_{x,\mu}$ must behave like an adjoint matter field defined at the site $x$ under the gauge transformation:
\begin{equation}
  X_{x,\mu} (=  U_{x,\mu}V_{x,\mu}^{-1} )
 \rightarrow \Omega_{x} X_{x,\mu} \Omega_{x}^{-1} = X_{x,\mu}^\prime
  , \quad \Omega_{x} \in G
  .
  \label{X-transf1}
\end{equation}
These properties of the decomposed variables under the gauge transformation are expected from the continuum version.

In what follows, we perform the lattice CFN decomposition in a constructive way. 
First, we consider the defining equation  which enables us to determine the decomposition uniquely. 
According to the  continuum formulation \cite{KSM08}, we introduce just a single color field $\bm{n}_{x}$ for  $G=SU(N)$ ($N \ge 2$).   This unit vector field is the initial or reference field to construct  possible other color fields which are necessary in the maximal case. 
We propose a lattice version of the first defining equation:
\textit{
The color field $\bm{n}_{x}$ is covariantly constant in the (matrix) background $V_{x,\mu}$:}
\begin{equation}
  \epsilon D_\mu^{(\epsilon)}[V]\bm{n}_{x} := V_{x,\mu} \bm{n}_{x+\mu} - \bm{n}_{x} V_{x,\mu} = 0 ,
  \label{lat-defeq-min1}
\end{equation}
where $D_\mu^{(\epsilon)}[V]$ is the lattice covariant derivative in the adjoint representation \cite{IKKMSS06}. 
This defining equation for the initial color field guarantees that all $N-1$ color fields $\bm{n}_{x}^{(k)}$ ($k=1, \cdots, N-1$) prepared in the maximal option are covariantly constant in the background $V_{x,\mu}$:
\begin{equation}
  \epsilon D_\mu^{(\epsilon)}[V]\bm{n}_{x}^{(k)} := V_{x,\mu} \bm{n}_{x+\mu}^{(k)} - \bm{n}_{x}^{(k)} V_{x,\mu} = 0 \quad (k=1, \cdots, N-1) .
  \label{lat-defeq-max1}
\end{equation}

The first defining equation  (\ref{lat-defeq-min1}) is just a lattice or group theoretical version of the Lie-algebra valued defining equations given already in the continuum formulation \cite{KSM08}. 
This choice is reasonable from the following observations.
i) When $V_{x,\mu} \equiv {\bf 1}$, the  covariant derivative $D_\mu^{(\epsilon)}[V]$ reduces to the (forward) lattice derivative 
$
 \partial_\mu^{(\epsilon)} \bm{n}_{x} := \epsilon^{-1}[\bm{n}_{x+\mu}-\bm{n}_{x}]
$.
ii) The  covariant derivative $D_\mu^{(\epsilon)}[V]$ reproduces correctly the continuum covariant derivative for the adjoint field in the naive continuum limit $\epsilon \rightarrow 0$  up to ${\cal O}(\epsilon)$:
\footnote{
By using the mid-point prescription, this is more improved up to ${\cal O}(\epsilon^2)$:
\begin{equation}
  \epsilon^{-1} [V_{x,\mu} {\bf n}_{x+\mu} - {\bf n}_{x} V_{x,\mu}] 
= \partial_\mu^{(\epsilon)} {\bf n}_{x^\prime} - i g[ \mathbf{V}_\mu(x^\prime), {\bf n}_{x^\prime}] 
-ig \epsilon/2 \{ \mathbf{V}_\mu(x^\prime) , \partial_\mu^{(\epsilon)} {\bf n}_{x^\prime} - i g[ \mathbf{V}_\mu(x^\prime), {\bf n}_{x^\prime}]  \} + {\cal O}(\epsilon^2) 
 .
\end{equation}
}
\begin{equation}
  \epsilon^{-1} [V_{x,\mu} {\bf n}_{x+\mu} - {\bf n}_{x} V_{x,\mu}] 
= \partial_\mu^{(\epsilon)} {\bf n}_{x} - i g[\mathbf{V}_\mu(x), {\bf n}_{x}] + {\cal O}(\epsilon) 
 .
\end{equation}
iii) The covariant derivative $D_\mu^{(\epsilon)}[V]$ obeys the correct transformation property, i.e., the adjoint rotation on a lattice: 
\begin{align}
  D_\mu^{(\epsilon)}[V] \bm{n}_{x} \rightarrow (D_\mu^{(\epsilon)}[V] \bm{n}_{x})^\prime = \Omega_{x}(D_\mu^{(\epsilon)}[V] \bm{n}_{x})\Omega_{x+\mu}^\dagger .
\end{align}
iv) The  first defining equation  is form-invariant under the gauge transformation, i.e.,
\begin{equation}
V_{x,\mu}^\prime \bm{n}_{x+\mu}^\prime =\bm{n}_{x}^\prime V_{x,\mu}^\prime  
 .
\end{equation}

Next, we give a general consideration to what extent the defining equations determine the decomposition uniquely, before proceeding to solving them explicitly. 
In order to consider the meaning of the second defining equation deeper, we return to the first equation.
It is important to observe that the decomposition (\ref{decomp-1}) is invariant under the simultaneous local transformation of  $V_{x,\mu}$ and $X_{x,\mu}$:
\footnote{
For another decomposition of the form:
$
 U_{x,\mu} = V_{x,\mu} X_{x,\mu} 
  ,
$
it is advantageous to take
\begin{equation}
  V_{x,\mu} \rightarrow V_{x,\mu} R_{x+\mu} , \quad 
  X_{x,\mu} \rightarrow R_{x+\mu}^{-1}  X_{x,\mu} , \quad 
R_{x+\mu} \in G 
  .
  \label{s-rotationb}
\end{equation}
}
\begin{equation}
 X_{x,\mu} \rightarrow X_{x,\mu} R_{x}^{-1} , \quad
 V_{x,\mu} \rightarrow R_{x} V_{x,\mu} , \quad 
R_{x} \in G 
  .
  \label{s-rotation}
\end{equation}
This is the extra   $G$ degrees of freedom which are absent in the original $G$ lattice theory written in terms of $U_{x,\mu}$.
In order to obtain the unique decomposition (\ref{decomp-1}), we must fix the extra degrees of freedom by imposing suitable conditions. 
Therefore, we examine to what extent the extra degrees of freedom (\ref{s-rotation}) is fixed by the first defining equation:
\begin{equation}
 \bm{n}_x V_{x,\mu} = V_{x,\mu} \bm{n}_{x+\mu} 
  .
  \label{def-1}
\end{equation}
It is obvious that the diagonal part related to the discrete symmetry of the center $Z(N)$ of $SU(N)$ is undetermined by the first defining equation:
\begin{equation}
  \exp (2\pi i n/N) \mathbf{1} \in Z(N) \quad (n=0,1,  \cdots , N) .
  \label{center}
\end{equation}
Suppose that the first defining equation holds after the local rotations (\ref{s-rotation}):
\begin{equation}
 \bm{n}_x R_{x} V_{x,\mu} = R_{x} V_{x,\mu} \bm{n}_{x+\mu} 
  .
\end{equation}
Combining this with the original equation (\ref{def-1}), we obtain the relationship
\begin{equation}
 \bm{n}_x R_{x} V_{x,\mu} = R_{x} \bm{n}_x V_{x,\mu}
 \Longleftrightarrow [\bm{n}_x , R_{x} ] V_{x,\mu} = 0
  .
\end{equation}
This implies that the degrees of freedom of $R_{x}$ satisfying the following equation cannot be determined by imposing the first defining equation alone. 
\begin{equation}
  [\bm{n}_x , R_{x} ] = 0
  .
  \label{G-dof}
\end{equation}

For the maximal option, the extra symmetry is $Z(N) \times H$, $H = U(1)^{N-1} \subset SU(N)$:
\begin{equation}
   R_{x} = \exp (2\pi i n/N) \exp \left\{ i \sum_{k=1}^{N-1} \alpha_{x}^{(k)} \bm{n}_{x}^{(k)} \right\} \in Z(N) \times H = Z(N) \times U(1)^{N-1} , 
   \label{extra-max}
\end{equation}
where $\alpha^{(k)} \in \mathbb{R}$ and $\{ \bm{n}_{x}^{(k)} \}$ is a maximal set of mutually commutable Hermitian generators for $U(1)^{N-1}$ with the traceless property ${\rm tr}(\bm{n}_{x}^{(k)})=0$ (see section 3).
In the minimal option, the extra symmetry is $Z(N) \times \tilde{H}$, $\tilde{H} = U(N-1) \subset SU(N)$:
\begin{equation}
   R_{x} = \exp (2\pi i n/N) \exp \left\{ i \alpha_{x} \bm{h}_{x}  \right\}
\exp \left\{ i \sum_{k=1}^{(N-1)^2-1} \beta_{x}^{(k)} \bm{u}_{x}^{(k)} \right\} \in Z(N) \times \tilde{H} = Z(N) \times U(N-1) ,
   \label{extra-min}
\end{equation}
where $\alpha_{x},  \beta_{x}^{(k)} \in \mathbb{R}$ and $\{ \bm{u}_{x}^{(k)} \}$ is a set of Hermitian generators of $SU(N-1)$ commutable with $\bm{h}_{x}:=\bm{n}_{x}^{(N-1)}$ (see section 4) with the traceless property ${\rm tr}(\bm{u}_{x}^{(k)})=0$.

Thus, we find that the degrees of freedom corresponding to $H$ or $\tilde{H}$ are left unfixed in the maximal or minimal options, respectively, even after solving the first defining equation.  
In order to obtain the unique decomposition, therefore, we must impose additional conditions to fix  degrees of freedom which remain undetermined by imposing the first defining equation. 
This role is played by the second defining equation.  
In the previous papers \cite{IKKMSS06,SKKMSI07,KSSMKI08}, we used as a second defining equation the condition:
\begin{equation}
  {\rm tr} [ X_{x,\mu} \bm{n}_{x}]  = 0 .
  \label{lat-defeq20}
\end{equation}
This is reasonable from the viewpoint of the naive continuum limit, since it leads to the second defining equation 
\begin{equation}
{\rm tr} [ \mathscr{X}_\mu(x)\bm{n}(x)]=0
  \label{defeq20}
\end{equation}
for the Lie-algebra valued field  $\mathscr{X}_\mu(x)$ in the continuum, as can be seen from 
\begin{align}
  {\rm tr} [ X_{x,\mu} \bm{n}_{x}]  
&= - ig \epsilon {\rm tr} [ \mathscr{X}_\mu(x)\bm{n}(x)] + O(\epsilon^2) 
 ,
\end{align}
using
$
 X_{x,\mu} =    \exp \left\{ -ig \epsilon \mathscr{X}_\mu(x) \right\} 
=  \mathbf{1} -ig \epsilon \mathscr{X}_\mu(x) + O(\epsilon^2) 
$.
In this paper, however, we are looking for a lattice version of the second defining equation valid for the Lie-group valued field  $X_{x,\mu}$, which is intrinsic for the lattice with arbitrary lattice spacing $\epsilon$. 
In the below, we will observe that (\ref{lat-defeq20}) is valid for $SU(2)$ exceptionally, but it is not valid for $SU(N)$, $N \ge 3$.
We need more care for the second defining equations.

Thus, imposing simultaneously the first and second defining equations uniquely fix the decomposition (\ref{decomp-1}) by eliminating the extra  gauge degrees of freedom associated to the decomposition.

\section{Maximal case}

For the Yang-Mills gauge theory with a gauge group $G=SU(N)$, it is convenient
 to introduce a set of  $(N^2-1)$-dimensional unit vector fields $ {\bf n}^{(k)}(x)$ 
($k=1, \cdots, r$) with the components $n_{(k)}^A(x)$, i.e., 
$
 \mathbf{n}^{(k)}(x) \cdot  \mathbf{n}^{(k)}(x) := n_{(k)}^A(x) n_{(k)}^A(x) = 1
$ 
($ A =1, 2, \dots, {\rm dim}G=N^2-1 $) 
where $r:={\rm rank}G=N-1$ is the rank  of the gauge group $G=SU(N)$.
But it is not essential to introduce $r$ fields $\bm{n}^{(k)}(x)$, since it is enough to introduce a single color field $\bm{n}(x)$, see \cite{KSM08}.
We omit the summation symbol for $A$ in what follows. 
The   $\bm n^{(k)}(x)$ fields having the value in the Lie algebra $\mathscr{G}$ are constructed according to 
\begin{equation}
 \bm n^{(k)}(x) = n_{(k)}^A(x) T^A  = U^\dagger(x) H_k U(x) 
 , \quad U(x) \in G
 ,
\end{equation}
where $H_k$ are generators in the Cartan subalgebra in the generators $T^A$ of the Lie-algebra $\mathscr{G}=su(N)$ of  $G=SU(N)$. 
We adopt  the normalization ${\rm tr}(T_A T_B)= \frac12 \delta_{AB}$.

It is known that an arbitrary complex-valued ($N$ by $N$) matrix $M$ can be decomposed into the following form:
\footnote{
The Lie-algebra version of this identity (\ref{eq:Mdecomp0}) was given in Appendix B of \cite{KSM08}.  This identity  (\ref{eq:Mdecomp0}) is obtained by the similar consideration, although we omit the derivation. 
} 
\begin{align}
M  &=M_{G/\tilde{H}}+M_{\tilde{H}} , 
\nonumber\\
 & M_{\tilde{H}} = \frac{1}{N} \mathrm{tr}(M) \mathbf{1} + 2\sum_{k=1}^{N-1} \mathrm{tr} (M\bm{n}^{(k)}) \bm{n}^{(k)} , 
\quad
   M_{G/\tilde{H}}    =\sum_{k=1}^{N-1}\left[  \bm{n}^{(k)} , \left[  \bm{n}^{(k)},M \right]  \right] .
\label{eq:Mdecomp0} 
\end{align}
The double commutator is calculated as
\begin{align}
& M_{G/\tilde{H}} := 
\sum_{k=1}^{N-1}
\left[  \bm{n}^{(k)}, \left[  \bm{n}^{(k)}, M \right]  \right]
=\left\{
\sum_{k=1}^{N-1} 
  \bm{n}^{(k)} \bm{n}^{(k)}, M \right\}  -%
\sum_{k=1}^{N-1}
2\bm{n}^{(k)}M\bm{n}^{(k)}\nonumber\\
&  =\left(  1-\frac{1}{N}\right)  M - 2\sum_{k=1}^{N-1}
\bm{n}^{(k)}M\bm{n}^{(k)} ,
\end{align}
by using the relation  
\begin{equation}
\sum_{k=1}^{N-1}
  \bm{n}^{(k)}   \bm{n}^{(k)}  = \frac12 \left(  1-\frac{1}{N}\right) \mathbf{1} ,
\end{equation}
which follows from
\begin{equation}
\bm{n}_{x}^{(k)}\bm{n}_{x}^{(l)}=\left\{
\begin{array}
[c]{lc}%
\frac{1}{\sqrt{2k(k+1)}}\bm{n}_{x}^{(l)} & (l<k)\\
\frac{1}{2N}\mathbf{1}-\frac{k-1}{\sqrt{2k(k+1)}}\bm{n}_{x}^{(k)}%
+\sum_{k+1 \le m \le N-1} \frac{1}{\sqrt{2m(m+1)}}\bm{n}_{x}^{(m)} & (l=k)\\
\frac{1}{\sqrt{2l(l+1)}}\bm{n}_{x}^{(k)} & (l>k)
\end{array}
\right.  .
\label{eq:nn-id}
\end{equation}

Therefore, the identity (\ref{eq:Mdecomp0}) is rewritten into%
\begin{equation}
M = \mathrm{tr}(M) \mathbf{1} + 2N \sum_{k=1}^{N-1} \mathrm{tr}(M\bm{n}^{(k)}) \bm{n}^{(k)}-2N 
\sum_{k=1}^{N-1}
\bm{n}^{(k)}M\bm{n}^{(k)} . 
\label{eq:Mdecomp1}%
\end{equation}

[Necessity: (\ref{lat-defeq-min1}) $\Longrightarrow$ (\ref{eq:X}) \& (\ref{eq:KXlc2})]
We apply the identity (\ref{eq:Mdecomp1}) to $X_{x,\mu}$ to obtain
\begin{equation}
X_{x,\mu}=\mathrm{tr}(X_{x,\mu})\mathbf{1}+2N%
\sum_{k=1}^{N-1}
\mathrm{tr}(X_{x,\mu}\bm{n}_{x}^{(k)})\bm{n}_{x}^{(k)}-2N%
\sum_{k=1}^{N-1}
\bm{n}_{x}^{(k)}X_{x,\mu}\bm{n}_{x}^{(k)}. 
\label{eq:X-identity}%
\end{equation}
By using $X_{x,\mu}=U_{x,\mu}V_{x,\mu}^{-1}$ and  $V_{x,\mu}^{-1}\bm{n}_{x}^{(k)}=\bm{n}_{x+\mu}^{(k)}V_{x,\mu}^{-1}$ which follows from the first defining equation,
the last term in eq.(\ref{eq:X-identity}) is cast into%
\begin{align}
&   
\bm{n}_{x}^{(k)}X_{x,\mu}\bm{n}_{x}^{(k)}
  = 
\bm{n}_{x}^{(k)}U_{x,\mu}V_{x,\mu}^{-1}\bm{n}_{x}^{(k)}
= 
\bm{n}_{x}^{(k)}U_{x,\mu}\bm{n}_{x+\mu}^{(k)}V_{x,\mu}^{-1}\nonumber\\
&  = 
\left(  \bm{n}_{x}^{(k)}U_{x,\mu}\bm{n}_{x+\mu}^{(k)}U_{x,\mu}^{-1} \right)  \left(  U_{x,\mu}V_{x,\mu}^{-1}\right)  
= 
\left(  \bm{n}_{x}^{(k)}U_{x,\mu}\bm{n}_{x+\mu}^{(k)}U_{x,\mu}^{-1} \right)  X_{x,\mu} .
\label{eq:Q1}%
\end{align}
By defining
\begin{equation}
K_{x,\mu} :=  \mathbf{1}+2N\sum_{k=1}^{N-1}\bm{n}_{x}^{(k)}U_{x,\mu} \bm{n}_{x+\mu}^{(k)}U_{x,\mu}^{-1} ,
\end{equation}
therefore, eq.(\ref{eq:X-identity}) is rewritten as
\begin{equation}
K_{x,\mu}  X_{x,\mu}
=\mathrm{tr}(X_{x,\mu})\mathbf{1}+2N \sum_{k=1}^{N-1}
\mathrm{tr}(X_{x,\mu}\bm{n}_{x}^{(k)})\bm{n}_{x}^{(k)} \quad (\text{no sum over $x,\mu$}) .
\label{eq:X-solve0}%
\end{equation}

Here we apply the polar decomposition theorem \cite{KSSMKI08} to $K_{x,\mu}$ which is assumed to be a regular matrix  (namely, the inverse $K_{x,\mu}^{-1}$ exists). 
Then we can obtain the unitary matrix $\hat{K}_{x,\mu}$ and a positive definite Hermitian matrix $H_{x,\mu}:=\sqrt{K_{x,\mu} K_{x,\mu}^\dagger}$ such that  
\begin{equation}
 K_{x,\mu} = \sqrt{K_{x,\mu} K_{x,\mu}^\dagger} \hat{K}_{x,\mu}  \Longleftrightarrow \hat{K}_{x,\mu}= (\sqrt{K_{x,\mu} K_{x,\mu}^\dagger})^{-1} K_{x,\mu}.
 \label{eq:pd}
\end{equation}

Applying (\ref{eq:pd}) to $K_{x,\mu}$ of (\ref{eq:X-solve0}),  eq.(\ref{eq:X-solve0}) reads
\begin{equation}
 \hat{K}_{x,\mu}  X_{x,\mu}
=\left( \sqrt{K_{x,\mu} K_{x,\mu}^\dagger} \right)^{-1}  \{
\mathrm{tr}%
(X_{x,\mu})\mathbf{1}+2N%
\sum_{k=1}^{N-1}
\mathrm{tr}(X_{x,\mu}\bm{n}_{x}^{(k)})\bm{n}_{x}^{(k)} \}
\quad (\text{no sum over $x,\mu$}) ,
\label{eq:KX}
\end{equation}
where $\hat{K}_{x,\mu}X_{x,\mu} \in U(N)$, since $X_{x,\mu} \in SU(N)$ and $\hat{K}_{x,\mu} \in U(N)$.

It is shown (see Appendix) that all $\bm{n}_{x}^{(k)}$ commute with $K_{x,\mu} K_{x,\mu}^\dagger$: 
\begin{equation}
 [ K_{x,\mu} K_{x,\mu}^\dagger , \bm{n}_{x}^{(k)}] = 0 \quad
( k=1, \cdots, N-1 ) .
\end{equation}
Then, using the same argument as that given in (2.19) and (A.7) in Appendix A of \cite{KSSMKI08}, it is also shown that 
\begin{equation}
 [ \left( \sqrt{K_{x,\mu} K_{x,\mu}^\dagger} \right)^{-1}, \bm{n}_{x}^{(k)}] = 0 \quad
( k=1, \cdots, N-1 ) .
\end{equation}
By applying the identity (\ref{eq:Mdecomp0}) to $M=\left( \sqrt{K_{x,\mu} K_{x,\mu}^\dagger} \right)^{-1}$, 
we find that $\left( \sqrt{K_{x,\mu} K_{x,\mu}^\dagger} \right)^{-1}$ is written as a linear combination of $\mathbf{1}$ and $\bm{n}_{x}^{(k)}$ ($k=1, \cdots, N-1$).
Thus the right-hand side of (\ref{eq:KX}) is written as a linear combination of $\mathbf{1}$ and $\bm{n}_{x}^{(k)}$with appropriate coefficients $a_{x}^{(0)}$ and $a_{x}^{(k)}$:  
\begin{equation}
 \hat{K}_{x,\mu}  X_{x,\mu}
= \exp \left\{ ia_{x}^{(0)} \textbf{1} + i \sum_{k=1}^{N-1} a_{x}^{(k)} \bm{n}_{x}^{(k)} \right\} ,
\label{eq:KXlc}
\end{equation}
since $\{ \textbf{1} , \bm{n}_{x}^{(1)}, ..., \bm{n}_{x}^{(N-1)} \}$ is a maximal set of mutually commuting matrices. 

Taking the determinant of both sides of (\ref{eq:KXlc}), we obtain for $\hat{K}_{x,\mu} \in U(N)$
$
 \det(\hat{K}_{x,\mu} )
= \det \left[ \exp \left\{ ia_{x}^{(0)} \textbf{1} \right\}  \right]  
$, 
where we have used $\det(X_{x,\mu})=1$ for $X_{x,\mu} \in SU(N)$ and  $\det   \left[ \exp  \left\{i  \sum_{k=1}^{N-1}  a_{x}^{(k)} \bm{n}_{x}^{(k)} \right\} \right]=1$, since 
$
 \det [ \exp  \left\{i  a_{x}^{(k)} \bm{n}_{x}^{(k)} \right\}  ]
=  \exp  \left\{i  a_{x}^{(k)} {\rm tr}(\bm{n}_{x}^{(k)}) \right\}  
= 1 
$
due to ${\rm tr}(\bm{n}_{x}^{(k)})=0$. Therefore, we obtain
\begin{equation}
\det (\hat{K}_{x,\mu})  
= (\exp \left\{ ia_{x}^{(0)}   \right\})^{N} ,
\end{equation}
and hence
\begin{equation}
\exp \left\{ ia_{x}^{(0)}  \textbf{1} \right\}
= e^{ 2\pi i p_{x}/N } \textbf{1} (\det (\hat{K}_{x,\mu}))^{1/N}  
\quad (p_{x}=0, \cdots, N-1) .
\end{equation}
Thus we have solved the first defining equation and the solution for $X_{x,\mu}$ is given by
\begin{align}
   X_{x,\mu}
=& \hat{K}_{x,\mu}^\dagger (\det (\hat{K}_{x,\mu}))^{1/N} 
 g_{x}^{-1} , 
\label{eq:X}
\end{align}
where
$
  \hat{K}_{x,\mu}= \left(\sqrt{K_{x,\mu} K_{x,\mu}^\dagger} \right)^{-1} K_{x,\mu}
$,
$
  \hat{K}_{x,\mu}^\dagger= K_{x,\mu}^\dagger \left(\sqrt{K_{x,\mu} K_{x,\mu}^\dagger}\right)^{-1} 
$
and
\begin{align}
 g_{x} = e^{ -2\pi i p_{x}/N }  \exp \left\{   -i \sum_{k=1}^{N-1} a_{x}^{(k)} \bm{n}_{x}^{(k)} \right\} \quad (p_{x}=0, \cdots, N-1) .
\end{align}
Indeed, $g_{x}$ is an element of the extra  symmetry considered in  (\ref{extra-max}): $Z(N) \times H$, $H = U(1)^{N-1} \subset SU(N)$.
Consequently, the solution of the first defining equation for $V_{x,\mu}=  X_{x,\mu}^\dagger U_{x,\mu}$ is given by 
\footnote{
If we introduce $\tilde{V}_{x,\mu}$ by
$
\tilde{V}_{x,\mu} := K_{x,\mu} U_{x,\mu} := U_{x,\mu} +2N\sum_{k=1}^{N-1}\bm{n}_{x}^{(k)}U_{x,\mu} \bm{n}_{x+\mu}^{(k)}  
$,
then $V_{x,\mu}$ is rewritten into the form given in \cite{KSSMKI08}: 
$
V_{x,\mu} 
=   g_{x} \left(\sqrt{\tilde{V}_{x,\mu} \tilde{V}_{x,\mu}^\dagger} \right)^{-1} \tilde{V}_{x,\mu} \left[ \det(\left(\sqrt{\tilde{V}_{x,\mu} \tilde{V}_{x,\mu}^\dagger}\right)^{-1} \tilde{V}_{x,\mu} ) \right]^{-1/N}  
   ,
$
where we have used 
$K_{x,\mu}U_{x,\mu}=\tilde{V}_{x,\mu}$,  $U_{x,\mu}^\dagger K_{x,\mu}^\dagger=\tilde{V}_{x,\mu}^\dagger$
and
$K_{x,\mu} K_{x,\mu}^\dagger=\tilde{V}_{x,\mu} \tilde{V}_{x,\mu}^\dagger$, and the resulting equality 
$\det(\hat{K}_{x,\mu})
=\det(\left(\sqrt{K_{x,\mu} K_{x,\mu}^\dagger}\right)^{-1} K_{x,\mu})
= \det(\left(\sqrt{\tilde{V}_{x,\mu} \tilde{V}_{x,\mu}^\dagger}\right)^{-1} \tilde{V}_{x,\mu} )
$.
}
\begin{align}
V_{x,\mu} 
= g_{x}  \hat{K}_{x,\mu} U_{x,\mu} (\det (\hat{K}_{x,\mu}))^{-1/N}   .
\label{eq:KXlc2}
\end{align}
It should be remarked that $g_{x}^{-1}$ (or $g_{x}$) in the above expressions for $X_{x,\mu}$ (or $V_{x,\mu}$) is undetermined from the first defining equation alone in agreement with the general consideration. %
%
%
In order to fix it, we must impose further conditions, i.e., the second defining equation,  by
equating  $g_{x}$ with an element $g_{x}^{0}$,
\begin{equation}
  g_{x}=g_{x}^{0} .
\end{equation}
 The simplest one is to take $g_{x}^{0}=\mathbf{1}$, or
\begin{equation}
  (\det (\hat{K}_{x,\mu}))^{-1/N}  \hat{K}_{x,\mu} X_{x,\mu}
= g_{x}^{0} =  \mathbf{1} .
\label{2nd}
\end{equation}

We can check that the naive continuum limit of (\ref{2nd}) reduces to the second defining equation (\ref{defeq20}) in the continuum formulation.
\footnote{
Note that $K_{x,\mu}=N \mathbf{1}+O(\epsilon)$, but $K_{x,\mu}K_{x,\mu}^\dagger=N \mathbf{1}+O(\epsilon^2)$ due to the cancellation of order $\epsilon$ terms and hence 
$(\sqrt{K_{x,\mu}K_{x,\mu}})^{-1}=(1/N)\mathbf{1}+O(\epsilon^2)$.
Moreover, 
$\det (\hat{K})=\det(1+O(\epsilon))=1+{\rm tr}(O(\epsilon))+ O(\epsilon^2)=1 + O(\epsilon^2)$ due to the tracelessness of the order $\epsilon$ terms by explicit calculation. Then $(\det (\hat{K}_{x,\mu}))^{-1/N}=1 + O(\epsilon^2)$. From (\ref{eq:X-solve0}), 
$K_{x,\mu} X_{x,\mu}=\mathrm{tr}(X_{x,\mu})\mathbf{1}+2N \sum_{k=1}^{N-1}
\mathrm{tr}[X_{x,\mu}\bm{n}_{x}^{(k)}]\bm{n}_{x}^{(k)}
=  \mathrm{tr}(\mathbf{1})\mathbf{1}+2N \sum_{k=1}^{N-1}
\mathrm{tr}[-i\epsilon g \mathscr{X}_{\mu}(x)\bm{n}_{x}^{(k)}]\bm{n}_{x}^{(k)}+ O(\epsilon^2) 
$.
Therefore, 
$\hat{K}_{x,\mu} X_{x,\mu}=   \mathbf{1}-2i\epsilon g   \sum_{k=1}^{N-1}
\mathrm{tr}[\mathscr{X}_{\mu}(x)\bm{n}_{x}^{(k)}]\bm{n}_{x}^{(k)}+ O(\epsilon^2)  
$. Thus, (\ref{2nd}) leads to $\mathrm{tr}[\mathscr{X}_{\mu}(x)\bm{n}_{x}^{(k)}]=0$ up to $O(\epsilon)$.} 

[Sufficiency: (\ref{eq:X}) \& (\ref{eq:KXlc2}) $\Longrightarrow$ (\ref{lat-defeq-min1})]
By using (\ref{nK=Km}):
\begin{equation}
 \bm{n}_{x}^{(k)} K_{x,\mu}
 = K_{x,\mu} U_{x,\mu} \bm{n}_{x+\mu}^{(k)}  U_{x,\mu}^\dagger ,
\end{equation}
we check that the above expression (\ref{eq:KXlc2}) for $V_{x,\mu}$ satisfies the first defining equation:
\begin{align}
\bm{n}_{x}^{(k)} V_{x,\mu}
 &= g_{x} (\det (\hat{K}_{x,\mu}))^{-1/N} \bm{n}^{(k)} \hat{K}_{x,\mu} U_{x,\mu}  
   \nonumber\\
 &=g_{x} (\det (\hat{K}_{x,\mu}))^{-1/N} (\sqrt{K_{x,\mu} K_{x,\mu}^\dagger})^{-1} \bm{n}_{x}^{(k)} K_{x,\mu} U_{x,\mu}  
   \nonumber\\
 &=g_{x} (\det (\hat{K}_{x,\mu}))^{-1/N} (\sqrt{K_{x,\mu} K_{x,\mu}^\dagger})^{-1} K_{x,\mu} U_{x,\mu} \bm{n}_{x+\mu}^{(k)} U_{x,\mu}^\dagger U_{x,\mu}  
   \nonumber\\
 &=g_{x}  (\det (\hat{K}_{x,\mu}))^{-1/N}  (\sqrt{K_{x,\mu} K_{x,\mu}^\dagger})^{-1} K_{x,\mu} U_{x,\mu} \bm{n}_{x+\mu}^{(k)} 
  =V_{x,\mu} \bm{n}_{x+\mu}^{(k)}  .
\end{align}
Thus, the sufficiency is shown irrespective of the extra part $g_{x}$.

\section{Minimal case}

Now we discuss the minimal case of $SU(N)$.
In the minimal case,  we introduce a single (traceless Hermitian) color  field: 
\begin{equation}
 \bm h(x)  \in G/\tilde{H} 
  .
\end{equation}

It is known that an arbitrary complex-valued ($N$ by $N$) matrix  $M$ can be decomposed into two parts:
\footnote{
The Lie-algebra version of this identity (\ref{eq:Mdecomp-002}) was given in Appendix B of \cite{KSM08}.  This identity (\ref{eq:Mdecomp-002}) is obtained by the similar consideration, although we omit the derivation. 
} 
\begin{align}
M  &=M_{G/\tilde{H}}+M_{\tilde{H}} , 
\nonumber\\
& M_{\tilde{H}} =\frac{1}{N}\mathrm{tr}\left( M \right) \mathbf{1}  +2\mathrm{tr }\left( M \bm{h} \right)  \bm{h} +\bar{M} , \quad
\bar{M}= \sum_{k=1}^{(N-1)^{2}-1} 2{\rm tr} \left(  M \bm{u}^{k}  \right) \bm{u}^{k} ,
\nonumber\\
& M_{G/\tilde{H}}    =2(1-1/N)\left[  \bm{h},\left[  \bm{h},M\right]  \right] ,
\label{eq:Mdecomp-002}%
\end{align}
where $M_{G/\tilde{H}}$ is the  $G/\tilde{H}$ part satisfying ${\rm tr}(M_{G/\tilde{H}} \bm{h})=0$ and $M_{\tilde{H}}$ is the  $\tilde{H}$ part satisfying $\left[\bm{h},M_{\tilde{H}} \right] =0$ where $\bar{M}$ is defined by a subset of generators   $\bm{u}^{k} \in su(N-1)$ such that $\left[ \bm{u}^{k} ,\bm{h} \right]  =0$ ($k=1,...,(N-1)^{2}-1$).

In order to make the following calculations easier, we adopt a specific representation:
\begin{equation}
 \bm h(x)=\bm n_r(x) :=U^\dagger(x) H_r U(x) \in G/\tilde{H} 
  ,
\end{equation}
where $H_r$ is the final Cartan matrix  given by
$
 H_r  
 = \frac{1}{\sqrt{2N(N-1)}} {\rm diag}(1,\cdots,1,-N+1) 
 .
$
The double commutator in
eq(\ref{eq:Mdecomp-002}) is calculated as 
\begin{align}
M_{G/\tilde{H}} &:= 2(1-1/N)\left[  \bm{h},\left[  \bm{h},M\right]  \right]
\nonumber\\
&  =2\left(  1-\frac{1}{N}\right)  \left(  \left\{  \bm{h} \bm{h} , M\right\}  -2\bm{h}M\bm{h}\right) 
\nonumber\\
&  =2\left(  1-\frac{1}{N}\right)  \left\{  \frac{1}{N}M-\frac{N-2}%
{\sqrt{2N(N-1)}}\left\{  \mathbf{h} , M\right\}  -2\bm{h}M\mathbf{h}%
\right\}  ,
\end{align}
where we have used  
\begin{equation}
\bm{h} \bm{h} =\frac{1}{2N}\mathbf{1-}\frac{N-2}{\sqrt{2N(N-1)}} \bm{h} .
\label{eq:Hidentity}%
\end{equation}
 Thus, the identity is cast into 
\begin{align}
&  \frac{N^{2}-2N+2}{N}M+\left(  N-2\right)  \sqrt{\frac{2(N-1)}{N}} 
\{ \bm{h}, M \}  +4\left(  N-1\right)  \bm{h}%
M\bm{h}\nonumber\\
&  =\mathrm{tr }\left(  M\right)  \mathbf{1}  +N\bar{M}+2N\mathrm{tr }\left(  M\bm{h}\right) \bm{h}.
\label{eq:min-id}
\end{align}

[Necessity: (\ref{lat-defeq-min1}) $\Longrightarrow$ (\ref{eq:Xmin}) \& (\ref{eq:KXlc2min})] 
Applying the identity (\ref{eq:min-id}) to $X_{x,\mu}$, we have
\begin{align}
&  \frac{N^{2}-2N+2}{N}X_{x,\mu}+\left(  N-2\right)  \sqrt{\frac{2(N-1)}{N} } \left\{  \bm{h}_{x}\mathbf{,}X_{x,\mu}\right\} + 4\left(
N-1\right)  \bm{h}_{x}X_{x,\mu}\bm{h}_{x}\nonumber\\
&  =\mathrm{tr}\left(  X_{x,\mu}\right)  \mathbf{1} 
+N\bar{X}_{x,\mu}+2N\mathrm{tr }\left(  X_{x,\mu
}\bm{h}_{x}\right)  \bm{h}_{x}. \label{eq:Xmin-000}%
\end{align}
By using  $X_{x,\mu}=U_{x,\mu}V_{x,\mu}^{-1}$ and  $V_{x,\mu}^{-1}\bm{h}_{x} =\bm{h}_{x+\mu} V_{x,\mu}^{-1}$ which follows from the first defining equation,
$\left\{  \mathbf{h,}X_{x,\mu}\right\} $ and $\bm{h}_{x}X_{x,\mu}\bm{h}_{x}$   are rewritten as
\begin{align}
\left\{  \bm{h}_{x}, X_{x,\mu} \right\}  &:= \bm{h}_{x}  X_{x,\mu}  + X_{x,\mu} \bm{h}_{x}   
= \left( \bm{h}_{x}+U_{x,\mu}\bm{h}_{x+\mu}U_{x,\mu}^{-1}\right)  X_{x,\mu} ,
\label{eq:XHmin-1}
\\
\bm{h}_{x}X_{x,\mu}\bm{h}_{x}  &  =\bm{h}_{x}U_{x,\mu}V_{x,\mu}^{-1} \bm{h}_{x}=\bm{h}_{x}U_{x,\mu}\bm{h}_{x+\mu}U_{x,\mu}^{-1}X_{x,\mu} .
\label{eq:XHmin-2}%
\end{align}
By defining
\begin{align}
L_{x,\mu}  
  =& \frac{N^{2}-2N+2}{N}\mathbf{1}+\left(  N-2\right)  \sqrt{\frac{2(N-1)}{N}%
}\left(  \bm{h}_{x}+U_{x,\mu}\bm{h}_{x+\mu}U_{x,\mu}^{-1}\right)
\nonumber\\
&
 +4\left(  N-1\right)  \bm{h}_{x}U_{x,\mu}\bm{h}_{x+\mu}U_{x,\mu}%
^{-1} ,
\end{align}
therefore,  we obtain
\begin{equation}
L_{x,\mu} X_{x,\mu}\mathbf{=}%
\mathrm{tr\mathrm{\mathrm{\mathrm{{}}}}}\left(  X_{x,\mu}\right)
\mathbf{1}+N\bar{X}_{x,\mu}+2N\mathrm{tr\mathrm{\mathrm{\mathrm{{}}}}}\left(
X_{x,\mu}\bm{h}_{x}\right)  \bm{h}_{x} .
\label{eq:XXmin-001}%
\end{equation}

We now apply the polar decomposition theorem to $L_{x,\mu}$ which is assumed to be a regular matrix  (namely, the inverse $L_{x,\mu}^{-1}$ exists). 
Then we can obtain the unitary matrix $\hat{L}_{x,\mu}$ and a positive definite Hermitian matrix $H_{x,\mu}:=\sqrt{L_{x,\mu}L_{x,\mu}^\dagger}$; 
\begin{equation}
 L_{x,\mu} = \sqrt{L_{x,\mu} L_{x,\mu}^\dagger} \hat{L}_{x,\mu} \Longleftrightarrow 
\hat{L}_{x,\mu}= (\sqrt{L_{x,\mu}L_{x,\mu} ^\dagger})^{-1} L_{x,\mu}.
\end{equation}
Then  (\ref{eq:XXmin-001}) reads 
\begin{equation}
 \hat{L}_{x,\mu}  X_{x,\mu}
=\left( \sqrt{L_{x,\mu} L_{x,\mu}^\dagger} \right)^{-1}  \{
\mathrm{tr\mathrm{\mathrm{\mathrm{{}}}}}\left(  X_{x,\mu}\right)
\mathbf{1}+N\bar{X}_{x,\mu}+2N\mathrm{tr\mathrm{\mathrm{\mathrm{{}}}}}\left(
X_{x,\mu}\bm{h}_{x}\right)  \bm{h}_{x} 
 \}
\quad (\text{no sum over $x,\mu$}) ,
\label{eq:KXmin}
\end{equation}
where $\hat{L}_{x,\mu}X_{x,\mu} \in U(N)$, since $X_{x,\mu} \in SU(N)$ and $\hat{L}_{x,\mu} \in U(N)$. 

It is shown that all $\bm{h}_{x}$ commute with $L_{x,\mu} L_{x,\mu}^\dagger$,  see Appendix: 
\begin{equation}
 [ L_{x,\mu} L_{x,\mu}^\dagger , \bm{h}_{x}] = 0 \quad
( k=1, \cdots, N-1 ) ,
\end{equation}
In the similar way to the maximal option, we can show  that   
\begin{equation}
 [ \left( \sqrt{L_{x,\mu} L_{x,\mu}^\dagger} \right)^{-1}, \bm{h}_{x} ] = 0  .
\end{equation}
By applying the identity (\ref{eq:Mdecomp-002}) to $M=\left( \sqrt{L_{x,\mu} L_{x,\mu}^\dagger} \right)^{-1}$, 
we find that $\left( \sqrt{L_{x,\mu} L_{x,\mu}^\dagger} \right)^{-1}$ is written as a linear combination of $\mathbf{1}$, $\bm{h}_{x}$ and $\bm{u}_{x}^{(\ell)}$ ($\ell=1, \cdots, (N-1)^2-1$).
Thus the right-hand side of (\ref{eq:KXmin}) is written as a linear combination of $\mathbf{1}$, $\bm{h}_{x}$ and $\bm{u}_{x}^{(\ell)}$  with appropriate coefficients $a_{x}^{(0)}$, $a_{x}$ and $a_{x}^{(\ell)}$:  
\begin{equation}
 \hat{L}_{x,\mu}  X_{x,\mu}
= \exp \left\{ ia_{x}^{(0)} \textbf{1} + ia_{x} \bm{h}_{x} + i \sum_{\ell=1}^{(N-1)^2-1} a_{x}^{(\ell)} \bm{u}_{x}^{(\ell)} \right\} ,
\label{eq:KXlcmin}
\end{equation}
since $\{ \textbf{1} , \bm{h}_{x}, \bm{u}_{x}^{(1)}, ..., \bm{u}_{x}^{((N-1)^2-1)} \}$ constitute the closed set of generators. 

Taking the determinant of both sides of (\ref{eq:KXlcmin}), we obtain
$
 \det(\hat{L}_{x,\mu} )
= \det \left[ \exp \left\{ ia_{x}^{(0)} \textbf{1} \right\}  \right] 
$,
where we have used $\det(X_{x,\mu})=1$ for $X_{x,\mu} \in SU(N)$, $\det \left[ \exp  \left\{i  \sum_{\ell=1}^{(N-1)^2-1} a_{x}^{(\ell)} \bm{u}_{x}^{(\ell)} \right\} \right]=1$, since $\exp  \left\{i  \sum_{\ell=1}^{(N-1)^2-1} a_{x}^{(\ell)} \bm{u}_{x}^{(\ell)} \right\} \in SU(N-1)$,
and $\det \left[ \exp \left\{ ia_{x}  \bm{h}_{x} \right\}  \right] =1$, since 
$
 \det [ \exp  \left\{i  a_{x}  \bm{h}_{x} \right\}  ]
=  \exp  \left\{i  a_{x}  {\rm tr}(\bm{h}_{x}) \right\}  
= 1 
$,
with  ${\rm tr}(\bm{h}_{x})=0$. 
Therefore, we obtain
\begin{equation}
\det (\hat{L}_{x,\mu})  
= (\exp \left\{ ia_{x}^{(0)}   \right\})^{N} ,
\end{equation}
and hence
\begin{equation}
\exp \left\{ ia_{x}^{(0)}  \textbf{1} \right\}
= e^{ 2\pi i q_{x}/N } \textbf{1} (\det (\hat{L}_{x,\mu}))^{1/N}  
\quad (q_{x}=0, \cdots, N-1) .
\end{equation}
Thus, we have solved the first defining equation and the solution for $X_{x,\mu}$ is given by
\begin{align}
   X_{x,\mu}
=& \hat{L}_{x,\mu}^\dagger (\det (\hat{L}_{x,\mu}))^{1/N} g_{x}^{-1} 
 ,
\label{eq:Xmin}
\end{align}
where
$
  \hat{L}_{x,\mu}= \left(1/\sqrt{L_{x,\mu} L_{x,\mu}^\dagger} \right)  L_{x,\mu}
$,
$
  \hat{L}_{x,\mu}^\dagger= L_{x,\mu}^\dagger \left(1/\sqrt{L_{x,\mu} L_{x,\mu}^\dagger}\right)  
$
and 
\begin{align}
   g_{x}
=e^{ -2\pi i q_{x}/N }  \exp \left\{  - ia_{x} \bm{h}_{x} - i \sum_{\ell=1}^{(N-1)^2-1} a_{x}^{(\ell)} \mathbf{u}_{x}^{(\ell)} \right\} \quad (q_{x}=0, \cdots, N-1) 
\end{align}
Indeed, $g_{x}$ is an element of the extra  symmetry considered in   (\ref{extra-min}): $Z(N) \times \tilde{H}$,  $\tilde{H} = U(N-1) \subset SU(N)$.
Consequently, the solution of the first defining equation for $V_{x,\mu}=X_{x,\mu}^\dagger U_{x,\mu}$ is given by
\begin{align}
V_{x,\mu} 
=  g_{x} \hat{L}_{x,\mu} U_{x,\mu} (\det (\hat{L}_{x,\mu}))^{-1/N}   
=   g_{x}  (\sqrt{L_{x,\mu} L_{x,\mu}^\dagger})^{-1} L_{x,\mu} U_{x,\mu} (\det (\hat{L}_{x,\mu}))^{-1/N}  
   .
\label{eq:KXlc2min}
\end{align}
Thus,  $g_{x}$  in the above expressions for $X_{x,\mu}$ and $V_{x,\mu}$ are undetermined from the first defining equation alone in agreement with the general consideration. 
In order to fix it, we must impose further conditions, i.e., the second defining equation.

[Sufficiency: (\ref{eq:Xmin}) \& (\ref{eq:KXlc2min}) $\Longrightarrow$  (\ref{lat-defeq-min1}) ]
By using (\ref{nL=Lm}):
\begin{equation}
 \bm{n}_{x}  L_{x,\mu}
 = L_{x,\mu} U_{x,\mu} \bm{n}_{x+\mu}   U_{x,\mu}^\dagger ,
\end{equation}
the sufficiency, i.e, the above expression (\ref{eq:KXlc2min}) for $V_{x,\mu}$ satisfies the first defining equation (\ref{lat-defeq-min1}) is shown in the same way as that in the maximal case.

\section{SU(2) case}

In the $SU(2)$ case, the maximal and minimal options are the same and cannot be distinguished. In fact, $K_{x,\mu}$ and $K_{x,\mu}$ are the same: 
\begin{equation}
K_{x,\mu} =  L_{x,\mu} = \mathbf{1}+ 4 \bm{n}_{x} U_{x,\mu} \bm{n}_{x+\mu} U_{x,\mu}^{-1}  ,
\quad
K_{x,\mu}^\dagger =  L_{x,\mu}^\dagger = \mathbf{1}+ 4  U_{x,\mu}  \bm{n}_{x+\mu}  U_{x,\mu}^\dagger  \bm{n}_{x} .
\end{equation}
The specific feature of the SU(2) case is that $K_{x,\mu} K_{x,\mu} ^\dagger$ is proportional to the unit matrix:
\begin{equation}
K_{x,\mu} K_{x,\mu} ^\dagger   = \frac12 {\rm tr}(K_{x,\mu} K_{x,\mu} ^\dagger) \mathbf{1}  ,
\end{equation}
which was already shown in the footnote 6 of \cite{IKKMSS06} where  $K_{x,\mu} K_{x,\mu} ^\dagger=\tilde{V}_{x,\mu} \tilde{V}_{x,\mu} ^\dagger$. 
Therefore, $\hat{K}_{x,\mu}$ is proportional to $K_{x,\mu}$, namely, $K_{x,\mu}$ agrees with the unitary $\hat{K}_{x,\mu}$ up to a numerical factor:
\begin{equation}
  \hat{K}_{x,\mu}= \left(\sqrt{K_{x,\mu} K_{x,\mu}^\dagger} \right)^{-1} K_{x,\mu}
=  \left(\sqrt{{\rm tr}(K_{x,\mu} K_{x,\mu} ^\dagger)/2}\right)^{-1} K_{x,\mu} .
\end{equation}
Therefore, we have 
\begin{align}
   X_{x,\mu}
=& \hat{K}_{x,\mu}^\dagger (\det (\hat{K}_{x,\mu}))^{1/2} 
 g_{x}^{-1} 
= \frac{(\det (K_{x,\mu}))^{1/2}}{{\rm tr}(K_{x,\mu} K_{x,\mu} ^\dagger)/2} K_{x,\mu} ^\dagger  g_{x}^{-1} , 
\end{align}
For SU(2), thus, the second defining equation (\ref{lat-defeq20}) is exceptionally satisfied when $g_{x}=\mathbf{1}$, since 
\begin{align}
  {\rm tr} [ X_{x,\mu} \bm{n}_{x}]  
=& \frac{(\det (K_{x,\mu}))^{1/2}}{{\rm tr}(K_{x,\mu} K_{x,\mu} ^\dagger)/2} {\rm tr} [ \bm{n}_{x} K_{x,\mu} ^\dagger  g_{x}^{-1}   ] 
\nonumber\\
=& \frac{(\det (K_{x,\mu}))^{1/2}}{{\rm tr}(K_{x,\mu} K_{x,\mu} ^\dagger)/2} {\rm tr} [ \bm{n}_{x} g_{x}^{-1}   +    U_{x,\mu}  \bm{n}_{x+\mu}  U_{x,\mu}^\dagger    g_{x}^{-1}  ]  
 ,
  \label{lat-defeq20r}
\end{align}
where we take into account ${\rm tr}(\bm{n}_{x})=0$ and the cyclicity of the trace.

\section{Conclusion}

In this paper, we have given a lattice version of the decomposition of the  Yang-Mills field,  which was originally proposed by Cho \cite{Cho80}, Faddeev and Niemi \cite{FN98}, and later developed by Kondo, Shinohara and Murakami \cite{KMS06,KSM08} in the continuum formulation. 
For the SU(N) gauge group, we have proposed a set of defining equations for specifying the decomposition of the gauge link variable and solve them  without using the  ansatz which was assumed in our own papers for $SU(2)$ and $SU(3)$.  
Finally, we have obtained the general form of the decomposition for $SU(N)$ gauge link variables, which confirms the previous results obtained for $SU(2)$ \cite{KKMSSI05,IKKMSS06,SKKMSI07} and $SU(3)$ \cite{KSSMKI08,SKKMSI07b}.

\appendix
\section{Proof of commutativity}
\subsection{Maximal case}

In what follows, we omit the indices, $x$ and $\mu$, and denote $\bm{n}_{x}$ by $\bm{n}$ and $\bm{n}_{x+\mu}$ by $\bm{n}^\prime$. 
An identity for the color field (\ref{eq:nn-id}) is 
\begin{equation}
\bm{n}^{(k)}\bm{n}^{(\ell)}
 =\frac1{2N}\delta_{k\ell}\bm1
  + \frac12 \sum_{m=1}^{N-1}\bar d_{k\ell m}\bm{n}^{(m)} ,
\end{equation}
where 
$
\bar d_{k \ell m}
 := 2{\rm Tr}(\{\bm{n}_{(k)},\bm{n}_{(\ell)}\}\bm{n}_{(m)}) 
$
and for $k < \ell < m$, 
$
\bar d_{kkk}
 =\frac{2(1-k)}{\sqrt{2k(k+1)}},
\quad
\bar d_{kkm}
 =\frac2{\sqrt{2m(m+1)}},
\quad
\bar d_{kmm}=0,
\quad
\bar d_{k\ell m}=0 .
$

For convenience, we  rewrite 
\begin{equation}
K=\bm1+2N\sum_{k=1}^{N-1}\bm{n}^{(k)}\bm{m}^{(k)}, \quad \bm{m}^{(k)}:=U\bm{n}^{(k)}{}^\prime U^\dagger .
\end{equation}
Then we have 
\begin{align}
\bm{n}_{(k)}K
 &=\bm{n}_{(k)}+2N\sum_{k=1}^{N-1}\bm{n}_{(k)}\bm{n}_{(\ell)}\bm{m}_{(\ell)}
   \nonumber\\
 &=\bm{n}_{(k)}
   +2N\sum_{k=1}^{N-1}
    \left[\frac1{2N}\delta_{k\ell}\bm1
          +\frac12\sum_{\ell=1}^{N-1}\bar d_{k\ell m}\bm{n}_{(m)}\right]\bm{m}_{(\ell)}
   \nonumber\\
 &=\bm{n}_{(k)}+\bm{m}_{(k)}
   +N\sum_{k=1}^{N-1}
     \sum_{\ell=1}^{N-1}\bar d_{k \ell m}\bm{n}_{(m)}\bm{m}_{(\ell)} .
\end{align}
On the other hand, we have
\begin{align}
K\bm{m}_{(k)}
 &=\bm{m}_{(k)}+2N\sum_{k=1}^{N-1}\bm{n}_{(\ell)}\bm{m}_{(\ell)}\bm{m}_{(k)}
   \nonumber\\
 &=\bm{m}_{(k)}
   +2N\sum_{k=1}^{N-1} \bm{n}_{(k)}
    \left[\frac1{2N}\delta_{k\ell}\bm1
          +\frac12\sum_{\ell=1}^{N-1}\bar d_{k\ell m}\bm{m}_{(m)}\right]
   \nonumber\\
 &=\bm{n}_{(k)}+\bm{m}_{(k)}
   +N\sum_{k=1}^{N-1}
     \sum_{\ell=1}^{N-1}\bar d_{k\ell m} \bm{n}_{(\ell)}\bm{m}_{(m)} .
\end{align}
Therefore, we have shown 
\begin{equation}
\bm{n}_{(k)}K=K\bm{m}_{(k)},
\quad
K^\dagger \bm{n}_{(k)}=\bm{m}_{(k)}K^\dagger .
\label{nK=Km}
\end{equation}
Thus, we conclude 
\begin{equation}
\bm{n}_{(k)}KK^\dagger
 =K\bm{m}_{(k)}K^\dagger
 =KK^\dagger \bm{n}_{(k)} 
 \Longleftrightarrow  
[\bm{n}_{(k)},KK^\dagger]=0 .
\end{equation}

\subsection{Minimal case}

Define $\bm{m}:=U\bm{n}^\prime U^\dagger$.
Then the formula (\ref{eq:Hidentity})
\begin{equation}
\bm{n}\bm{n}=\frac1{2N}\bm1-\frac{N-2}{\sqrt{2N(N-1)}}\bm{n},
\label{nn}
\end{equation}
leads to 
\begin{equation}
\bm{m}\bm{m}=\frac1{2N}\bm1-\frac{N-2}{\sqrt{2N(N-1)}}\bm{m} .
\label{mm}
\end{equation}
Then we have
\begin{equation}
\bm{n}\bm{n}-\bm{m}\bm{m}=-\frac{N-2}{\sqrt{2N(N-1)}}(\bm{n}-\bm{m}) .
\label{nn-mm}
\end{equation}
Moreover, we find from (\ref{nn}) and (\ref{mm})
\begin{equation}
\bm{n}\bm{n}\bm{m}-\bm{n}\bm{m}\bm{m}=-\frac1{2N}(\bm{n}-\bm{m}).
\label{nnm-nmm}
\end{equation}
We rewrite $L$ in the form
\begin{equation}
L=\frac{N^2-2N+2}N\bm 1
  +(N-2)\sqrt{\frac{2(N-1)}N}(\bm{n}+\bm{m})
  +4(N-1)\bm{n}\bm{m},
\end{equation}
to calculate $ \bm{n}L-L\bm{m}$ by using (\ref{nn-mm}) and (\ref{nnm-nmm}) 
\begin{align}
& \bm{n}L-L\bm{m}
   \nonumber\\
 &=\frac{N^2-2N+2}N(\bm{n}-\bm{m})
   +(N-2)\sqrt{\frac{2(N-1)}N}(\bm{n}\bm{n}-\bm{m}\bm{m})
   +4(N-1)(\bm{n}\bm{n}\bm{m}-\bm{n}\bm{m}\bm{m})
   \nonumber\\
 &=\left[
   \frac{N^2-2N+2}N
   -\frac{(N-2)^2}N
   -\frac{2(N-1)}N
   \right](\bm{n}-\bm{m})
=0 .
\label{nL=Lm}
\end{align}
In the similar way to the maximal case, thus, we conclude 
\begin{equation}
\bm{n} LL^\dagger
 =L\bm{m} L^\dagger
 =LL^\dagger \bm{n} 
 \Longleftrightarrow  
[\bm{n} ,LL^\dagger]=0 .
\end{equation}

\section*{Acknowledgments}
K.-I. K. is grateful to High Energy Physics Theory Group and Theoretical Hadron Physics Group in the University of Tokyo, especially, Prof. Tetsuo Hatsuda for kind hospitality extended to him on sabbatical leave.
This work is financially supported by Grant-in-Aid for Scientific Research (C) 21540256 from Japan Society for the Promotion of Science
(JSPS).


\end{document}